\documentclass[10pt]{article}

\usepackage[T1]{fontenc}
\usepackage[utf8]{inputenc}
\usepackage[a4paper,top=0.72in,bottom=0.75in,left=0.65in,right=0.65in]{geometry}
\usepackage{microtype}
\usepackage[numbers,sort&compress]{natbib}

\usepackage{amsmath,amssymb}
\usepackage{graphicx}
\usepackage{booktabs}
\usepackage{adjustbox}
\usepackage{multirow}
\usepackage{xcolor}
\usepackage{colortbl}
\usepackage{algorithm}
\usepackage{algorithmicx}
\usepackage{algpseudocode}
\usepackage{hyperref}
\usepackage{balance}

\definecolor{purple}{RGB}{128,0,128}
\hypersetup{
  colorlinks=true,
  linkcolor=blue,
  citecolor=blue,
  urlcolor=blue,
  pdftitle={FSDBN: Foreground-Aware EEG-Visual Alignment via Dynamic Brain Networks},
  pdfauthor={Yiheng Liu, Chuhang Zheng, Peiliang Gong, Jingtao Liu, Daoqiang Zhang, and Qi Zhu}
}

\title{FSDBN: Foreground-Aware EEG-Visual Alignment via Dynamic Brain Networks}

\author{%
Yiheng Liu$^{1,*}$, Chuhang Zheng$^{2,*}$, Peiliang Gong$^{1}$,
Jingtao Liu$^{1}$, Daoqiang Zhang$^{1}$, and Qi Zhu$^{1,\dagger}$\\[0.6em]
\small $^{1}$College of Artificial Intelligence,
Nanjing University of Aeronautics and Astronautics, Nanjing, China\\
\small $^{2}$The International Joint Institute of Tianjin University, Fuzhou,
Tianjin University, Tianjin, China\\[0.4em]
\scriptsize
\href{mailto:sx2524003@nuaa.edu.cn}{\texttt{sx2524003@nuaa.edu.cn}}
\enspace
\href{mailto:chzheng@tju.edu.cn}{\texttt{chzheng@tju.edu.cn}}
\enspace
\href{mailto:plgong@nuaa.edu.cn}{\texttt{plgong@nuaa.edu.cn}}
\enspace
\href{mailto:liujingtao@nuaa.edu.cn}{\texttt{liujingtao@nuaa.edu.cn}}
\enspace
\href{mailto:dqzhang@nuaa.edu.cn}{\texttt{dqzhang@nuaa.edu.cn}}
\enspace
\href{mailto:zhuqi@nuaa.edu.cn}{\texttt{zhuqi@nuaa.edu.cn}}
}

\date{}

\begin{document}

\twocolumn[
\begin{@twocolumnfalse}
\maketitle
\vspace{-1em}
\end{@twocolumnfalse}
]

\begingroup
\renewcommand{\thefootnote}{\fnsymbol{footnote}}
\footnotetext[1]{Yiheng Liu and Chuhang Zheng contributed equally to this work.}
\footnotetext[2]{Corresponding author}
\endgroup

\begin{abstract}
EEG-based visual decoding provides a non-invasive pathway to interpret visual semantics. However, existing methods ignore the perceptual asymmetry between foreground and background in complex scenes, leading to background interference and semantic misalignment. Compounding this challenge, EEG signals exhibit rapid temporal dynamics and non-stationary spatial patterns, posing a challenge for capturing the time-varying brain connectivity activated by focal visual attention. To address these limitations, we propose \textbf{FSDBN}, a unified framework for robust EEG-visual decoding. Specifically, we introduce a \textbf{Semantic-Consistent Saliency Alignment (SCSA)} module to disentangle semantic foregrounds from background noise under the saliency-semantic joint constraints. Furthermore, we devise a \textbf{Semantic-Prior Dynamic Gating Foreground Fusion} to adaptively regulate the contributions of foreground and background features for better representation learning. Correspondingly, we model EEG signals as \textbf{adaptive spatiotemporal brain networks}, where functional connectivity dynamically reorganizes to capture neural responses to salient foregrounds, facilitating robust alignment between dynamic neural representations and the visual structure. Our method achieves a top-1 accuracy of \textbf{69.0\%} and a top-5 accuracy of \textbf{92.2\%} on the zero-shot brain-to-image retrieval task, surpassing previous state-of-the-art methods. Code is available at \url{https://github.com/LiuYiheng1/FSDBN-EEG}.
\end{abstract}

\vspace{0.8em}

\section{Introduction}

EEG-based visual decoding seeks to reconstruct or recognize visual content directly from neural activity, offering a non-invasive pathway to understanding visual cognition and enabling practical brain–computer interface applications \citep{benchetrit2023brain,song2024decoding,wang2024cbramod}.
Leveraging its high temporal resolution and portability, EEG serves as a vital tool for capturing the dynamic neural processes underlying visual perception \citep{jiang2024labram,wang2024cbramod,yang2023biot}.
Recent advances in deep learning and cross-modal representation learning have further promoted EEG-based decoding in tasks such as image classification, cross-modal retrieval, and 3D visual reconstruction \citep{ma2025brainclip,wang2024decoding,guo2025neuro3d}, highlighting its potential as an accessible and temporally precise modal for visual semantic analysis \citep{benchetrit2023brain,li2025brainflora}.
\begin{figure}[t]
\centering
\includegraphics[width=\linewidth]{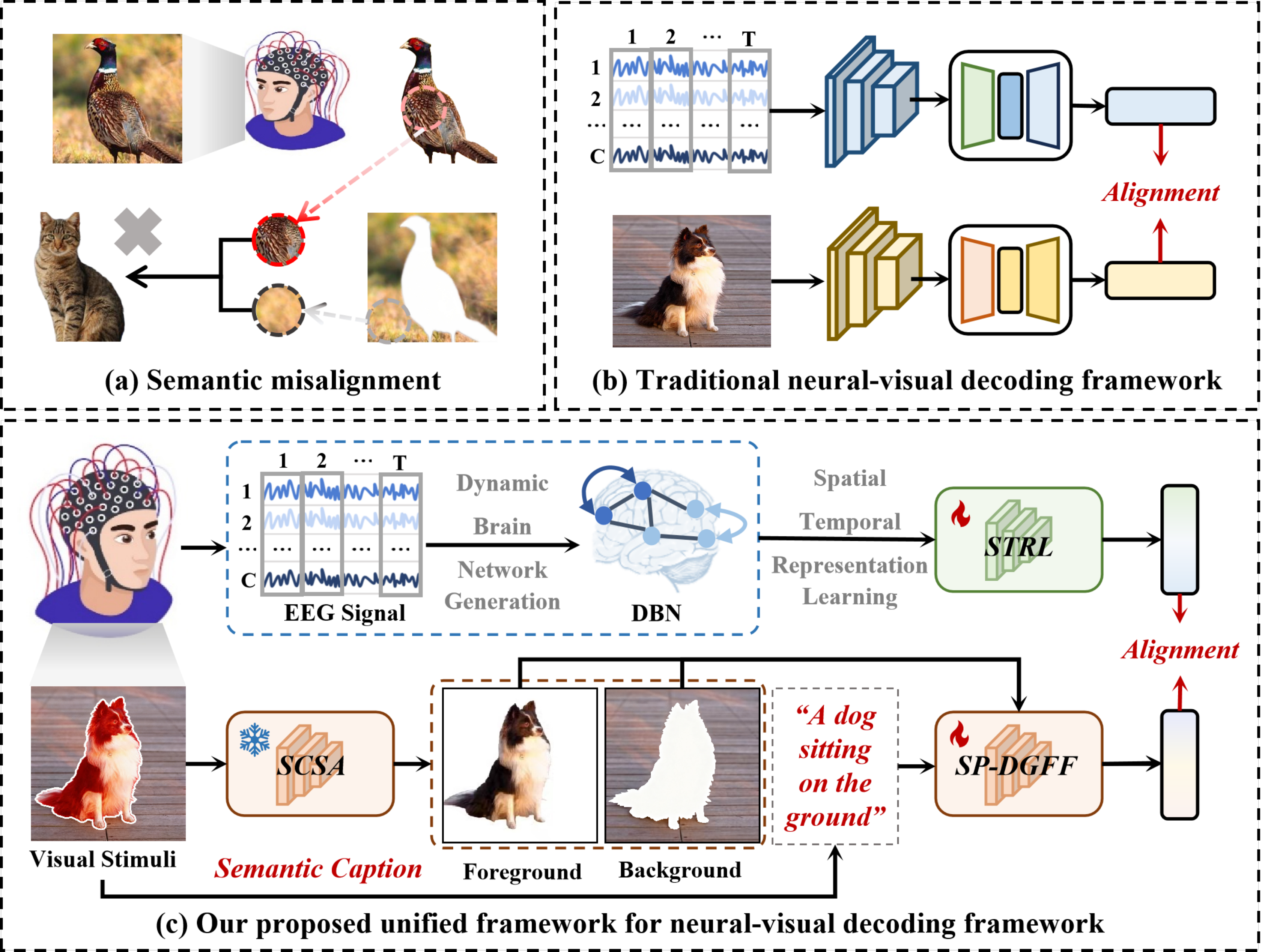}
\caption{Motivation and framework overview of FSDBN. (a) Semantic misalignment problem. (b) Traditional neural-visual decoding framework. (c) Our proposed unified framework.}
\label{fig:dongji}

\end{figure}
Conventional approaches typically align EEG representations with global visual features extracted from entire images \citep{song2024decoding,lu2023minddiffuser}. By leveraging convolutional neural networks or vision transformers and employing contrastive learning or regression objectives \citep{ozcelik2023natural,ferrante2024decoding,zhang2025cognitioncapturer}, these methods achieve reasonable performance on simple classification tasks. However, these methods implicitly assume a uniform contribution of all image regions to semantic perception \citep{li2024visual}, failing to distinguish task-relevant foregrounds from background noise. Consequently, decoding performance is often degraded by irrelevant background information in complex scenes \citep{ruotsalo2023feeling,de2024perceptual}. As illustrated in Figure~\ref{fig:dongji}(a), when a subject views a complex scene containing a pheasant, traditional global feature-based approaches struggle to disentangle the semantic foreground (the bird) from background clutter (e.g., grassland), leading to semantic misalignment where neural signals may be incorrectly associated with background regions.

To mitigate background interference, prior studies have introduced visual attention or saliency mechanisms to reweight image features \citep{wang2021salient,liu2024vst++,zhu2024separate}. However, most of these approaches rely on low-level saliency cues or implicit attention weights \citep{chen2024collaborative}, lacking explicit constraints on semantic consistency. Consequently, the attended regions often fail to align with true semantic entities \citep{li2025neuraldiffuser}, thereby undermining the stability and interpretability of cross-modal alignment.

Other approaches seek to incorporate fine-grained semantic information or brain network structures by leveraging pretrained semantic models \citep{radford2021learning,devlin2019bert} or graph-based EEG representations \citep{li2024multi,jin2024pgcn,sun2025spatial}. While these efforts improve discriminability, they typically treat semantic information as static features and model brain networks as static graphs \citep{liu2024self}, failing to capture the dynamic interplay between semantic modulation and neural network reorganization that characterizes visual perception \citep{bassett2017network}. Recent multi-subject decoding approaches attempt to address individual variability but still lack explicit foreground modeling \citep{yin2025mind}.

From a neurocognitive perspective, human visual semantic perception is neither uniform nor passive. In contrast, it is highly selective and hierarchically organized \citep{felleman1991distributed}: the visual system preferentially prioritizes task-relevant semantic foregrounds \citep{peelen2014attention}, while large-scale brain networks are dynamically reconfigured under top-down semantic modulation \citep{corbetta2002control,bastos2012canonical} to integrate information efficiently. This asymmetric processing of foreground and background regions serves as a fundamental mechanism enabling effective interpretation of complex visual scenes \citep{oliva2006building}. Figure~\ref{fig:dongji}(b-c) illustrates the traditional neural-visual decoding framework and our proposed approach. Unlike existing methods (b) that align global visual features with EEG representations without explicit foreground differentiation, our framework (c) incorporates three brain-inspired components: (1) Semantic-Consistent Saliency Alignment (SCSA) for explicit foreground-background decoupling, (2) Semantic-Prior Dynamic Gating Foreground Fusion (SP-DGFF) for adaptive top-down modulation, and (3) Dynamic Brain Network-driven EEG encoding to capture time-varying neural interactions.

Motivated by these observations, we identify a key limitation of existing EEG-based visual decoding methods: the absence of explicit modeling of semantic foreground prioritization and dynamic brain network reorganization induced by visual semantic perception \citep{van2014attention,bressler2010large}. To address these challenges, we propose a Foreground-Semantic Dynamic Brain Network (FSDBN) for EEG visual decoding. By computationally integrating semantic-consistent foreground modeling with dynamic brain network analysis, FSDBN effectively emulates the key mechanisms underlying visual semantic perception \citep{friston2011functional,park2018dynamic}. Our approach draws inspiration from the latest advances in autoregressive visual decoding \citep{dai2026autoregressive} and multi-modal neural encoding \citep{li2025brainflora}, while addressing their limitations in foreground-background disentanglement.

Our main contributions are summarized as follows:
\begin{itemize}
    \item We propose a brain-inspired \textbf{ Semantic-Consistent Saliency Alignment (SCSA)} mechanism to explicitly model the \emph{selective nature of human visual perception}. By disentangling task-relevant semantic foregrounds from background clutter, the proposed framework emulates the asymmetric foreground–background processing of the human visual system and alleviates semantic misalignment in EEG-based visual decoding.
    \item We introduce a \textbf{Semantic-Prior Dynamic Gating Foreground Fusion} mechanism that \emph{simulates top-down semantic modulation in the brain}. Instead of treating semantics as static auxiliary information, high-level semantic representations are formulated as adaptive feedback signals that dynamically regulate visual encoding, reflecting semantic-driven feedback processing in human vision.
    \item To model the \emph{dynamic neural reorganization underlying visual semantic perception}, we represent EEG signals as \textbf{Adaptive Spatiotemporal Brain Networks}. This formulation captures time-varying functional connectivity driven by focused visual attention and enables biologically plausible alignment between neural dynamics and saliency-guided visual representations.
    \item Extensive experiments on multiple EEG–visual decoding benchmarks demonstrate consistent and substantial performance improvements over state-of-the-art methods (e.g., \textbf{+18.1\% Top-1 accuracy on THINGS-EEG}), validating the effectiveness of incorporating neurocognitive priors for brain-inspired visual decoding.
\end{itemize}

\section{Related Work}
\subsection{Visual Neural Decoding}

EEG--visual decoding aims to establish mappings between neural signals and visual stimuli, and has become a central topic in brain decoding research. Early and recent studies predominantly rely on deep neural networks to model temporal EEG representations and align them with visual features through regression or contrastive objectives, achieving notable progress in visual classification and cross-modal retrieval tasks \citep{song2024decoding,ma2025brainclip,zhang2025cognitioncapturer,dai2026autoregressive,liu2025vieeg}. However, most existing approaches extract global visual representations from entire images, implicitly assuming uniform semantic contributions across spatial regions. This assumption is frequently violated in complex natural scenes, where semantic perception is dominated by a small number of task-relevant foreground entities, leading to performance degradation due to background interference.

To alleviate this issue, several works introduce visual attention, saliency mechanisms, or spatial degradation strategies to suppress background noise \citep{wang2021salient,liu2024vst++,zhu2024separate,li2025neuraldiffuser}. While such methods partially improve robustness, they often rely on low-level saliency cues or static attention weights without enforcing semantic consistency, resulting in attended regions that may diverge from true semantic entities. More recent approaches further explore spatial transformations, such as adaptive blurring or retinotopic mapping, to reduce neural--pixel granularity mismatch \citep{lu2023minddiffuser,de2024perceptual}.

With the emergence of large-scale pretrained vision--language models, recent studies incorporate fine-grained semantic embeddings to enhance EEG--visual alignment \citep{radford2021learning,li2025brainflora,wang2024cbramod}. Nevertheless, semantic information is typically treated as static auxiliary features, overlooking the top-down semantic modulation and dynamic functional reorganization observed in human visual cognition \citep{erouglu2025topdown,bastos2012canonical}. In contrast, FSDBN explicitly models semantic foreground prioritization and dynamic brain network reorganization within a unified framework, enabling brain-inspired visual decoding without resorting to spatial degradation or rigid architectural decomposition.

\subsection{Dynamic Brain Networks}

Graph-based methods have emerged as effective tools for modeling structured interactions among brain regions, significantly enhancing EEG representation capacity across diverse cognitive tasks \citep{li2024multi,jin2024pgcn,sun2025spatial}. Nevertheless, most existing EEG--visual decoding approaches rely on static connectivity graphs with fixed adjacency structures, which fundamentally conflict with the rapid and task-dependent functional reorganization observed during visual perception \citep{calhoun2014chronnectome,vidaurre2017brain}. By assuming temporally invariant connectivity, these methods implicitly treat all time intervals as functionally equivalent, thereby diluting task-relevant connectivity patterns induced by shifts in visual attention and semantic processing.

Several recent studies attempt to incorporate temporal dynamics through recurrent architectures or sliding-window-based graph modeling \citep{ding2025dgat,li2025multichannel}. While these approaches capture coarse temporal variations, they typically decouple network dynamics from semantic learning objectives, treating connectivity evolution as auxiliary or post-hoc descriptors rather than integral components of cross-modal alignment. As a result, they fail to model the synergistic interplay between time-varying neural interactions and visual semantic representations that characterizes natural perception. In contrast, our dynamic brain network modeling jointly learns time-varying functional connectivity and semantic alignment, enabling neural representations to evolve in correspondence with foreground-driven visual semantics while maintaining computational efficiency and topological consistency.

\section{Method}
\subsection{Overview}
\begin{figure*}[t]
  \centering
  \includegraphics[width=\textwidth]{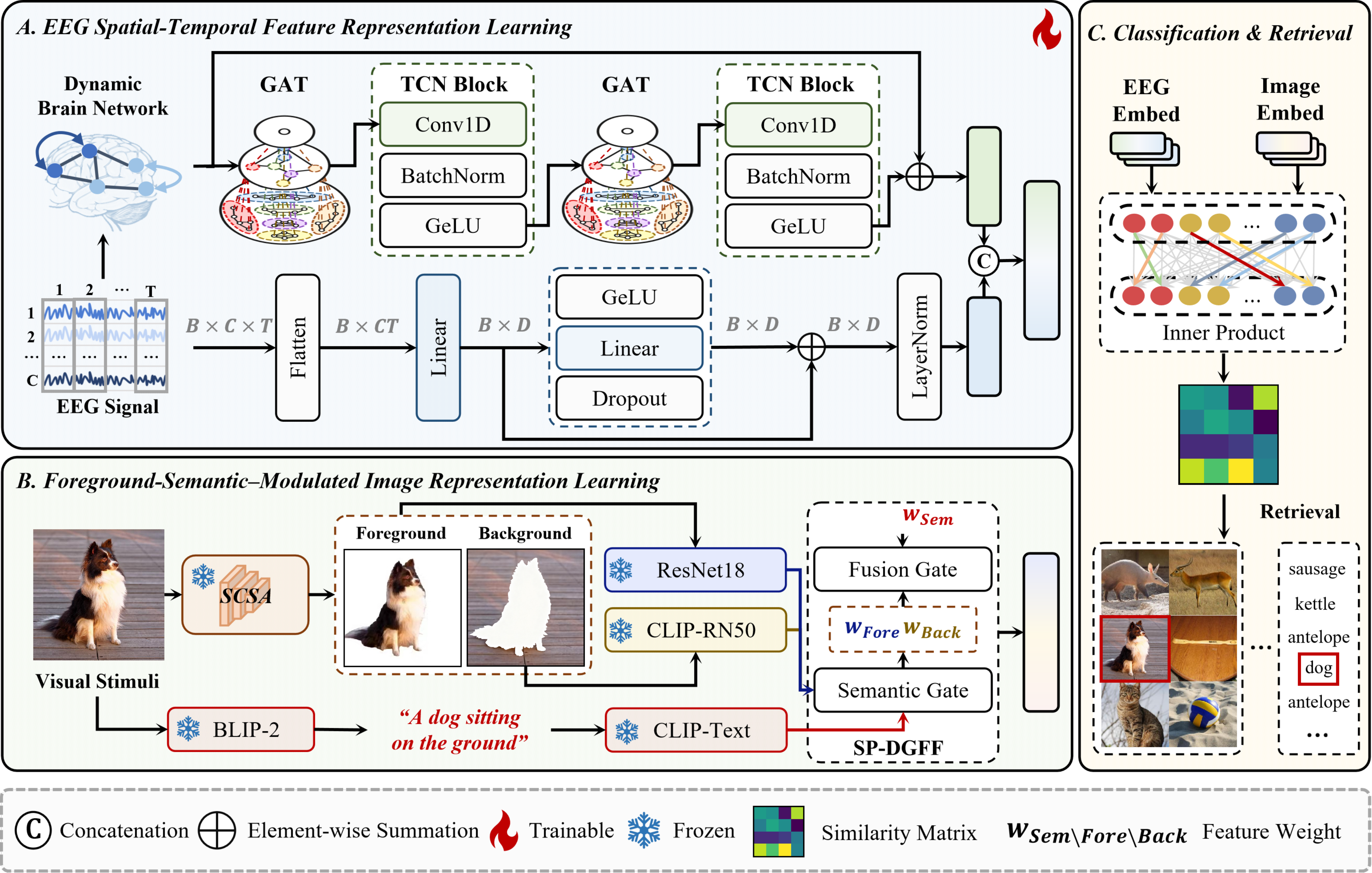}
  \caption{
  Overall framework of FSDBN.
  (A) \textbf{EEG Representation Learning}: Raw EEG signals are modeled as dynamic brain networks, alternating between Graph Attention Networks (GAT) and Temporal Convolutional Networks (TCN) to capture spatio-temporal dynamics.
  (B) \textbf{Visual Representation Learning}: Input images are decomposed into semantic foreground and background regions via Semantic-Consistent Saliency Alignment (SCSA). BLIP-2 extracts global semantic descriptions, while ResNet18 and CLIP encode visual features. The Semantic-Prior Dynamic Gating Foreground Fusion module adaptively integrates foreground and background features under semantic guidance.
  (C) \textbf{Cross-modal Alignment}: EEG and visual embeddings are aligned in a shared embedding space to support zero-shot retrieval task.
  Red flames indicate trainable parameters; blue snowflakes indicate frozen pre-trained models.
  }
  \label{fig:method}

\end{figure*}
We propose FSDBN, a brain-inspired framework for EEG–visual representation alignment that explicitly models semantic foreground prioritization and dynamic neural organization. The design of FSDBN is grounded in neurocognitive evidence indicating that visual perception is characterized by asymmetric foreground–background processing and top-down semantic modulation over large-scale brain networks \citep{bressler2010large,van2014attention}. As illustrated in Figure~\ref{fig:method}, FSDBN comprises three tightly integrated components: (i) semantic-consistent foreground–background decomposition, (ii) Semantic-Prior Dynamic Gating Foreground Fusion, and (iii) dynamic brain network-based EEG modeling.

\subsection{Semantic-Consistent Saliency Alignment}
\begin{figure}[t]
\vskip 0.2in
\begin{center}
\centerline{\includegraphics[width=\linewidth]{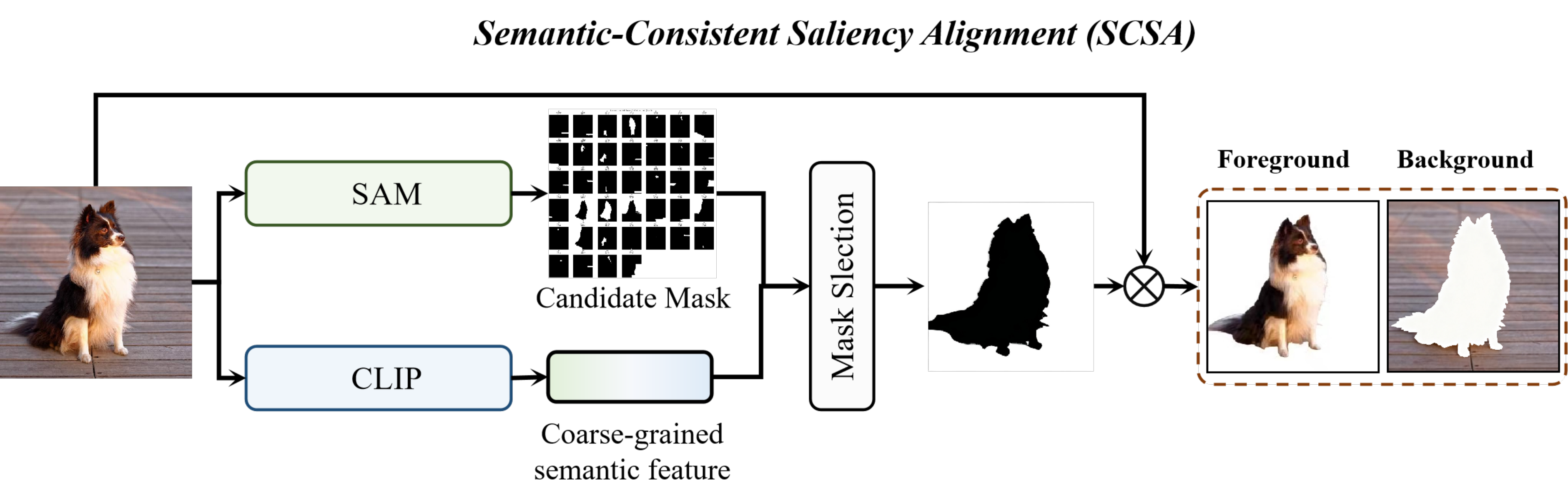}}
\caption{Pipeline of the proposed Semantic-Consistent Saliency Alignment (SCSA) module.}
\label{fig:SCSA}

\end{center}
\end{figure}
Rather than relying solely on bottom-up saliency, we model visual perception as a process jointly influenced by bottom-up cues and top-down semantic signals \citep{kriegeskorte2015deep,lotte2007review,bar2006topdown}. As illustrated in Figure~\ref{fig:SCSA}, the proposed SCSA module integrates saliency estimation with semantic guidance to produce semantically consistent foreground representations, thereby reducing the impact of background interference.

Given an input image $\mathbf{I}$, a saliency detector produces a map $\mathbf{M} \in [0,1]^N$, where $M_i$ denotes the importance of the $i$-th spatial region. A visual encoder extracts region-level features $\mathbf{F} \in \mathbb{R}^{N \times d}$, and a global semantic embedding $\mathbf{s} \in \mathbb{R}^d$ is obtained from a pretrained vision--language model. The saliency map provides an initial estimation of candidate foreground regions, while the semantic embedding serves as a global reference that constrains their semantic interpretation.

To enforce alignment between saliency-weighted visual features and the semantic representation, we define a semantic consistency loss:
\begin{equation}
\mathcal{L}_{sc} = 1 - \cos\left(
\mathbf{s},
\frac{1}{\sum_{i=1}^{N} M_i}
\sum_{i=1}^{N} M_i \mathbf{F}_i
\right),
\end{equation}
which encourages the aggregated salient features to be consistent with the global semantic concept. By optimizing this objective, the model is guided to focus on regions that are not only visually salient but also semantically relevant, while suppressing distractive background responses.

The overall training objective is defined as
\begin{equation}
\mathcal{L} = \mathcal{L}_{contrastive} + \lambda \mathcal{L}_{sc},
\end{equation}
where $\mathcal{L}_{contrastive}$ aligns EEG and visual embeddings through a contrastive learning paradigm. Specifically, we adopt an InfoNCE formulation:
\begin{equation}
\mathcal{L}_{contrastive} = - \frac{1}{B} \sum_{i=1}^{B} 
\log \frac{\exp\left(\mathrm{sim}(\mathbf{z}_i^e, \mathbf{z}_i^v)/\tau\right)}
{\sum_{j=1}^{B} \exp\left(\mathrm{sim}(\mathbf{z}_i^e, \mathbf{z}_j^v)/\tau\right)},
\end{equation}
where $\mathbf{z}_i^e$ and $\mathbf{z}_i^v$ denote the EEG and visual embeddings of the $i$-th sample, $\mathrm{sim}(\cdot,\cdot)$ is cosine similarity, $\tau$ is a temperature parameter, and $B$ is the batch size.

Although SCSA introduces no additional learnable parameters, $\mathcal{L}_{sc}$ provides effective supervision by influencing the optimization of visual representations. Gradients from $\mathcal{L}_{sc}$ are back-propagated through the saliency-weighted aggregation, directly updating the region-level features $\mathbf{F}$. As a result, semantically consistent regions are progressively emphasized, while inconsistent regions are suppressed. This mechanism can be interpreted as an implicit form of top-down modulation, where high-level semantic information reshapes low-level feature responses.

Finally, based on the refined saliency map, foreground and background features are explicitly disentangled as
\begin{equation}
\mathbf{F}^{fg} = \mathbf{M} \odot \mathbf{F}, \quad
\mathbf{F}^{bg} = (1 - \mathbf{M}) \odot \mathbf{F},
\end{equation}
yielding structured representations that are both robust to background noise and aligned with semantic content. This explicit decomposition further enhances the discriminability of visual features for downstream EEG--visual alignment.

\subsection{Semantic-Prior Dynamic Gating Foreground Fusion}
Neurocognitive studies suggest that semantic knowledge exerts top-down modulation over early visual representations, dynamically regulating the contribution of different perceptual components during visual processing \citep{van2014attention}. Inspired by this mechanism, we model semantic information as an explicit modulation signal rather than static auxiliary features.

Given the global semantic embedding $\mathbf{s}$, we compute adaptive gating coefficients for foreground and background features via
\begin{equation}
[\alpha^{fg}, \alpha^{bg}] = \mathrm{Softmax}(\mathbf{W}_g \mathbf{s}),
\end{equation}
where $\mathbf{W}_g \in \mathbb{R}^{2 \times d}$ is a learnable projection matrix, and $\alpha^{fg}$ and $\alpha^{bg}$ represent the semantic-driven importance weights assigned to foreground and background features, respectively.

To obtain compact visual representations, spatial pooling is applied to foreground and background features:
\begin{equation}
\mathbf{f}^{fg} = \mathrm{Pool}(\mathbf{F}^{fg}), \quad
\mathbf{f}^{bg} = \mathrm{Pool}(\mathbf{F}^{bg}),
\end{equation}
where $\mathrm{Pool}(\cdot)$ denotes global average pooling over spatial locations.

The final visual embedding is then computed as a semantic-driven fusion:
\begin{equation}
\mathbf{z}^v =
\alpha^{fg} \cdot \mathbf{f}^{fg} +
\alpha^{bg} \cdot \mathbf{f}^{bg}.
\end{equation}
This formulation allows semantic information to dynamically regulate the relative contributions of foreground and background visual features, enabling adaptive and context-sensitive visual representation learning.

\subsection{Dynamic Brain Network Modeling for EEG}

\begin{algorithm}[t]
\caption{Dynamic Brain Network Modeling for EEG Encoding}
\label{alg:dbn}
\begin{algorithmic}[1]
\Require 
    Multichannel EEG signal $\mathbf{X}^e \in \mathbb{R}^{C \times T}$; 
    Number of channels $C$; 
    Time steps $T$; 
    Embedding dimension $d_e$;
    GAT layers $\mathcal{G}$; 
    TCN layers $\mathcal{T}$
\Ensure 
    EEG embedding $\mathbf{z}^e \in \mathbb{R}^{d_e}$

\State // Step 1: Temporal segmentation and node initialization
\For{$t = 1$ to $T$}
    \State $\mathbf{H}_t \leftarrow \text{LinearProjection}(\mathbf{X}^e_{:,t})$ \Comment{Project to $d_e$ dims}
    \State $\mathbf{H}_t \in \mathbb{R}^{C \times d_e}$ \Comment{Node features at time $t$}
\EndFor

\State // Step 2: Dynamic spatial modeling via GAT
\State Initialize empty list $\mathcal{H}' \leftarrow []$
\For{$t = 1$ to $T$}
    \State $\mathbf{H}'_t \leftarrow \text{GAT}(\mathbf{H}_t; \mathcal{G})$ 
    \Comment{Adaptive channel interactions}
    \State Append $\mathbf{H}'_t$ to $\mathcal{H}'$
\EndFor

\State // Step 3: Temporal dependency modeling via TCN
\State $\mathbf{Z}^e \leftarrow \text{TCN}(\mathcal{H}'; \mathcal{T})$ 
\Comment{$\mathbf{Z}^e \in \mathbb{R}^{C \times T' \times d_e}$}

\State // Step 4: Global temporal pooling
\State $\mathbf{z}^e \leftarrow \text{TemporalPooling}(\mathbf{Z}^e)$ 
\Comment{Aggregate over time and channels}

\State \Return $\mathbf{z}^e$
\end{algorithmic}
\end{algorithm}

EEG signals exhibit pronounced non-stationary spatiotemporal dynamics, reflecting time-varying functional connectivity patterns induced by cognitive and perceptual processes~\cite{hutchison2013dynamic}. To capture this property, we model EEG signals as dynamic brain networks whose topology and node interactions evolve over time. The detailed encoding procedure is presented in Algorithm~\ref{alg:dbn}, and the mathematical formulation is described below.

Given multichannel EEG recordings $\mathbf{X}^e \in \mathbb{R}^{C \times T}$, where $C$ denotes the number of EEG channels and $T$ denotes the number of temporal samples, we first project the raw signal into a $d_e$-dimensional feature space. Specifically, at each time step $t \in \{1, \dots, T\}$, the node feature matrix $\mathbf{H}_t \in \mathbb{R}^{C \times d_e}$ is computed via a learnable linear projection:
\begin{equation}
    \mathbf{H}_t = \mathbf{X}^e_{:,t} \mathbf{W}_p^\top,
    \label{eq:node_init}
\end{equation}
where $\mathbf{X}^e_{:,t} \in \mathbb{R}^{C \times 1}$ represents the vector of voltage readings across all $C$ channels at time $t$, and $\mathbf{W}_p \in \mathbb{R}^{d_e \times 1}$ denotes the projection weight matrix. This formulation treats each EEG channel as a graph node, allowing node representations to evolve over time and capture instantaneous neural states.

To model dynamic functional interactions between channels, we employ Graph Attention Networks (GAT) to update node features at each time step. Formally, for each time slice $t$, the spatial refinement is performed as:
\begin{equation}
    \mathbf{H}'_t = \operatorname{GAT}(\mathbf{H}_t; \boldsymbol{\theta}_{\text{gat}}) \in \mathbb{R}^{C \times d_e},
    \label{eq:gat}
\end{equation}
where $\mathbf{H}'_t$ denotes the updated node representations at time $t$, and $\boldsymbol{\theta}_{\text{gat}}$ represents the learnable parameters of the GAT layers (including attention weights and transformation matrices). The attention mechanism adaptively estimates connectivity strength between channels based on their feature similarity, enabling data-driven modeling of time-varying functional connectivity without predefined adjacency matrices.

Following the spatial refinement, we capture temporal dependencies across the sequence of graph-enhanced features. The sequence $\{\mathbf{H}'_t\}_{t=1}^{T}$ is processed by a Temporal Convolutional Network (TCN) with kernel size $k$ and dilation factors $\{d_l\}_{l=1}^{L}$:
\begin{equation}
    \mathbf{Z}^e = \operatorname{TCN}\left(\{\mathbf{H}'_t\}_{t=1}^{T}; \boldsymbol{\theta}_{\text{tcn}}\right) \in \mathbb{R}^{C \times T' \times d_e},
    \label{eq:tcn}
\end{equation}
where $\mathbf{Z}^e$ represents the spatiotemporal EEG feature tensor, $T'$ denotes the temporal resolution after convolution (with $T' < T$ due to valid padding), and $\boldsymbol{\theta}_{\text{tcn}}$ denotes the TCN parameters. The TCN aggregates long-range temporal context while preserving the causal ordering of neural events, which is critical for modeling the evolution of perceptual processes.

Finally, to obtain a compact representation suitable for cross-modal alignment, we apply temporal pooling over the spatiotemporal volume $\mathbf{Z}^e$. The global EEG embedding $\mathbf{z}^e \in \mathbb{R}^{d_e}$ is computed as:
\begin{equation}
    \mathbf{z}^e = \operatorname{TemporalPool}\left(\mathbf{Z}^e\right) = \frac{1}{C \cdot T'} \sum_{c=1}^{C} \sum_{\tau=1}^{T'} \mathbf{Z}^e_{c,\tau,:},
    \label{eq:pooling}
\end{equation}
where $\mathbf{Z}^e_{c,\tau,:} \in \mathbb{R}^{d_e}$ denotes the feature vector at channel $c$ and time $\tau$, and the pooling operation averages over both spatial (channel) and temporal dimensions to produce a trial-level summary. This embedding $\mathbf{z}^e$ is subsequently aligned with the semantic representations via the contrastive objective.

\section{Experiments and Results}

\subsection{Datasets and Implementation Details}

\paragraph{THINGS-EEG} THINGS-EEG \cite{gifford2022large} is a large-scale EEG visual perception dataset consisting of recordings from 10 subjects under the Rapid Serial Visual Presentation (RSVP) paradigm. The training set contains 1,654 semantic concepts, each associated with 10 images, with each image repeated 4 times per subject. The test set includes 200 concepts, each with a single image repeated 80 times per subject to ensure high signal-to-noise ratio (SNR). We follow the preprocessing pipeline described in prior work \cite{spampinato2017deep}, including band-pass filtering (0.1--100 Hz), artifact removal, and baseline correction. EEG trials corresponding to repeated stimuli are averaged, resulting in 16,540 training samples and 200 test samples per subject.

\paragraph{THINGS-MEG} THINGS-MEG \cite{hebart2023things} includes MEG recordings from 4 subjects with 271 channels. The training set consists of 1,854 concepts with 12 images per concept and one repetition per image, while the test set includes 200 concepts with one image repeated 12 times. The same preprocessing pipeline is adopted, and repeated trials are averaged to enhance SNR.

\paragraph{Implementation Details.} Our model is implemented in PyTorch. The visual branch uses SAM~\cite{kirillov2023segment} (ViT-H) for automatic mask generation, CLIP~\cite{radford2021learning} (ViT-B/16) for semantic features, and BLIP-2~\cite{li2023blip} for caption generation. The EEG encoder adopts a 3-layer GAT (8 heads) and a 3-layer TCN (kernel size 3), with embedding dimension 256. Training is conducted for 100 epochs using Adam ($10^{-4}$, batch size 64), and the contrastive temperature $\tau$ is set to 0.07.

\subsection{Comparison with Baselines}

\begin{table*}[t]
  \caption{Top-1 and Top-5 accuracy (\%) for 200-way zero-shot retrieval on THINGS-EEG}
  \centering
  \adjustbox{max width=\linewidth}{
  \begin{tabular}{lccccccccccccccccccccccc}
    \toprule
    \multirow{2}{*}{Method} & \multicolumn{2}{c}{Subject 1} & \multicolumn{2}{c}{Subject 2} & \multicolumn{2}{c}{Subject 3} & \multicolumn{2}{c}{Subject 4} & \multicolumn{2}{c}{Subject 5} & \multicolumn{2}{c}{Subject 6} & \multicolumn{2}{c}{Subject 7} & \multicolumn{2}{c}{Subject 8} & \multicolumn{2}{c}{Subject 9} & \multicolumn{2}{c}{Subject 10} & \multicolumn{2}{c}{Avg} \\
    \cmidrule(lr){2-3} \cmidrule(lr){4-5} \cmidrule(lr){6-7} \cmidrule(lr){8-9} \cmidrule(lr){10-11} \cmidrule(lr){12-13} \cmidrule(lr){14-15} \cmidrule(lr){16-17} \cmidrule(lr){18-19} \cmidrule(lr){20-21} \cmidrule(lr){22-23}
    & top-1 & top-5 & top-1 & top-5 & top-1 & top-5 & top-1 & top-5 & top-1 & top-5 & top-1 & top-5 & top-1 & top-5 & top-1 & top-5 & top-1 & top-5 & top-1 & top-5 & top-1 & top-5 \\
    \midrule
    \multicolumn{23}{c}{\textbf{Intra-subject : train and test on one subject}} \\
    \midrule
    BraVL~\cite{du2023decoding} & 6.1 & 17.9 & 4.9 & 14.9 & 5.6 & 17.4 & 5.0 & 15.1 & 4.0 & 13.4 & 6.0 & 18.2 & 6.5 & 20.4 & 8.8 & 23.7 & 4.3 & 14.0 & 7.0 & 19.7 & 5.8 & 17.5 \\
    NICE~\cite{song2024decoding} & 13.2 & 39.5 & 13.5 & 40.3 & 14.5 & 42.7 & 20.6 & 52.7 & 10.1 & 31.5 & 16.5 & 44.0 & 17.0 & 42.1 & 22.9 & 56.1 & 15.4 & 41.6 & 17.4 & 45.8 & 16.1 & 43.6 \\
    NICE-SA~\cite{song2024decoding} & 13.3 & 40.2 & 12.1 & 36.1 & 15.3 & 39.6 & 15.9 & 49.0 & 9.8 & 34.4 & 14.2 & 42.4 & 17.9 & 43.6 & 18.2 & 50.2 & 14.4 & 38.7 & 16.0 & 42.8 & 14.7 & 41.7 \\
    NICE-GA~\cite{song2024decoding} & 15.2 & 40.1 & 13.9 & 40.1 & 14.7 & 42.7 & 17.6 & 48.9 & 9.0 & 29.7 & 16.4 & 44.4 & 14.9 & 43.1 & 20.3 & 52.1 & 14.1 & 39.7 & 19.6 & 46.7 & 15.6 & 42.8 \\
    MB2C~\cite{wei2024mb2c} & 23.6 & 56.3 & 22.6 & 50.5 & 26.3 & 60.1 & 34.8 & 67.0 & 21.3 & 53.0 & 31.0 & 62.3 & 25.0 & 54.8 & 39.0 & 69.3 & 27.5 & 59.3 & 33.1 & 70.8 & 28.4 & 60.3 \\
    ATM-S~\cite{li2024visual} & 25.6 & 60.4 & 22.0 & 54.5 & 25.0 & 62.4 & 31.4 & 60.9 & 12.9 & 43.0 & 21.3 & 51.1 & 30.5 & 61.5 & 38.8 & 72.0 & 24.4 & 51.5 & 29.1 & 63.5 & 26.1 & 58.1 \\
    CogCap~\cite{zhang2025cognitioncapturer} & 27.2 & 59.5 & 28.7 & 57.0 & 37.2 & 66.1 & 37.7 & 63.2 & 21.8 & 47.8 & 31.6 & 58.1 & 32.8 & 59.6 & 47.6 & 73.5 & 33.4 & 57.7 & 35.1 & 63.6 & 33.3 & 60.6 \\
    Neural-MCRL~\cite{li2025neural} & 27.5 & 64.0 & 28.5 & 61.5 & 37.0 & 69.0 & 35.0 & 66.0 & 22.5 & 51.5 & 31.5 & 61.0 & 31.5 & 62.5 & 42.0 & 74.5 & 30.5 & 59.5 & 37.5 & 71.0 & 32.4 & 64.1 \\
    VE-SDN~\cite{chen2024visual} & 32.6 & 63.7 & 34.4 & 69.9 & 38.7 & 73.5 & 39.8 & 72.0 & 29.4 & 58.6 & 34.5 & 68.8 & 34.5 & 68.3 & 49.3 & 79.8 & 39.0 & 69.6 & 39.8 & 75.3 & 37.2 & 69.9 \\
    UBP~\cite{wu2025bridging} & \underline{41.2} & \underline{70.5} & \underline{51.2} & \underline{80.9} & \underline{51.2} & \underline{82.0} & \underline{51.1} & \underline{76.9} & \underline{42.2} & \underline{72.8} & \underline{57.5} & \underline{83.5} & \underline{49.0} & \underline{79.9} & \underline{58.6} & \underline{85.8} & \underline{45.1} & \underline{76.2} & \underline{61.5} & \underline{88.2} & \underline{50.9} & \underline{79.7} \\
    \rowcolor{purple!20}
    \textbf{Ours} & \textbf{54.0} & \textbf{87.0} & \textbf{67.5} & \textbf{92.5} & \textbf{71.0} & \textbf{95.5} & \textbf{57.0} & \textbf{81.5} & \textbf{53.0} & \textbf{84.0} & \textbf{71.5} & \textbf{93.5} & \textbf{78.0} & \textbf{98.5} & \textbf{88.0} & \textbf{99.0} & \textbf{66.0} & \textbf{91.5} & \textbf{84.0} & \textbf{98.5} & \textbf{69.0} & \textbf{92.2} \\
    \midrule
    \multicolumn{23}{c}{\textbf{Inter-subject : leave one subject out for test}} \\
    \midrule
    BraVL~\cite{du2023decoding} & 2.3 & 8.0 & 1.5 & 6.3 & 1.4 & 5.9 & 1.7 & 6.7 & 1.5 & 5.6 & 1.8 & 7.2 & 2.1 & 8.1 & 2.2 & 7.6 & 1.6 & 6.4 & 2.3 & 8.5 & 1.8 & 7.0 \\
    NICE~\cite{song2024decoding} & 7.6 & 22.8 & 5.9 & 20.5 & 6.0 & 22.3 & 6.3 & 20.7 & 4.4 & 18.3 & 5.6 & 22.2 & 5.6 & 19.7 & 6.3 & 22.0 & 5.7 & 17.6 & 8.4 & 28.3 & 6.2 & 21.4 \\
    NICE-SA~\cite{song2024decoding} & 7.0 & 22.6 & 6.6 & 23.2 & 7.5 & 23.7 & 5.4 & 21.4 & 6.4 & 22.2 & 7.5 & 22.5 & 3.8 & 19.1 & 8.5 & 24.4 & 7.4 & 22.3 & 9.8 & 29.6 & 7.0 & 23.1 \\
    NICE-GA~\cite{song2024decoding} & 5.9 & 21.4 & 6.4 & 22.7 & 5.5 & 20.1 & 6.1 & 21.0 & 4.7 & 19.5 & 6.2 & 22.5 & 5.9 & 19.1 & 7.3 & 25.3 & 4.8 & 18.3 & 6.2 & 26.3 & 5.9 & 21.6 \\
ATM-S~\cite{li2024visual}
& 10.5 & 26.8
& 7.2 & 24.8
& \textbf{\cellcolor{purple!20}11.9}
& \textbf{\cellcolor{purple!20}33.8}
& \textbf{\cellcolor{purple!20}14.7}
& \textbf{\cellcolor{purple!20}39.4}
& 7.0 & 23.9
& 11.1 & \underline{35.8}
& \underline{16.1} & \underline{43.5}
& \textbf{\cellcolor{purple!20}15.0}
& \textbf{\cellcolor{purple!20}40.3}
& 4.9 & 22.7
& \textbf{\cellcolor{purple!20}20.5}
& \textbf{\cellcolor{purple!20}46.5}
& 11.8 & \underline{33.7} \\

UBP~\cite{wu2025bridging}
& \underline{11.5} & \underline{29.7}
& \underline{15.5} & \underline{40.0}
& 9.8 & 27.0
& \underline{13.0} & \underline{32.3}
& \underline{8.8} & \textbf{\cellcolor{purple!20}33.8}
& \underline{11.7} & 31.0
& 10.2 & 23.8
& \underline{12.2} & \underline{32.2}
& \textbf{\cellcolor{purple!20}15.5}
& \textbf{\cellcolor{purple!20}40.5}
& 16.0 & \underline{43.5}
& \underline{12.4} & 33.4 \\

\textbf{\cellcolor{purple!20}Ours}
& \textbf{\cellcolor{purple!20}20.0}
& \textbf{\cellcolor{purple!20}46.0}
& \textbf{\cellcolor{purple!20}23.0}
& \textbf{\cellcolor{purple!20}41.0}
& \underline{10.5} & \underline{31.5}
& 12.5 & 30.0
& \textbf{\cellcolor{purple!20}9.0}
& \underline{28.5}
& \textbf{\cellcolor{purple!20}12.5}
& \textbf{\cellcolor{purple!20}37.5}
& \textbf{\cellcolor{purple!20}19.0}
& \textbf{\cellcolor{purple!20}47.5}
& 11.5 & 29.5
& \underline{13.5} & \underline{38.5}
& \underline{18.9} & 41.0
& \textbf{\cellcolor{purple!20}15.0}
& \textbf{\cellcolor{purple!20}37.1} \\
    \bottomrule
  \end{tabular}
  }
  \label{tab:things-eeg}
\end{table*}

\begin{table}[t]
\caption{Top-1 and Top-5 accuracy (\%) for 200-way zero-shot retrieval on THINGS-MEG.}
\centering
\adjustbox{max width=\linewidth}{
\begin{tabular}{lcccccccccc}
    \toprule
    \multirow{2}{*}{Method} & \multicolumn{2}{c}{Subject 1} & \multicolumn{2}{c}{Subject 2} & \multicolumn{2}{c}{Subject 3} & \multicolumn{2}{c}{Subject 4} & \multicolumn{2}{c}{Avg} \\
    \cmidrule(lr){2-3} \cmidrule(lr){4-5} \cmidrule(lr){6-7} \cmidrule(lr){8-9} \cmidrule(lr){10-11}
    & top-1 & top-5 & top-1 & top-5 & top-1 & top-5 & top-1 & top-5 & top-1 & top-5 \\
    \midrule
    \multicolumn{11}{c}{\textbf{Intra-subject: train and test on one subject}} \\
    \midrule
    NICE~\cite{song2024decoding} & 9.6 & 27.8 & 18.5 & 47.8 & 14.2 & 41.6 & 9.0 & 26.6 & 12.8 & 36.0 \\
    NICE-SA~\cite{song2024decoding} & 9.8 & 27.8 & 18.6 & 46.4 & 10.5 & 38.4 & 11.7 & 27.2 & 12.7 & 35.0 \\
    NICE-GA~\cite{song2024decoding} & 8.7 & 30.5 & 21.8 & 56.6 & 16.5 & 49.7 & 10.3 & 32.3 & 14.3 & 42.3 \\
    UBP~\cite{wu2025bridging} & \underline{15.0} & \underline{38.0} & \underline{46.0} & \underline{80.5} & \underline{27.3} & \underline{59.0} & \underline{18.5} & \underline{43.5} & \underline{26.7} & \underline{55.2} \\
    \rowcolor{purple!20}
    \textbf{Ours} & \textbf{20.4} & \textbf{46.8} & \textbf{53.7} & \textbf{98.5} & \textbf{42.9} & \textbf{85.8} & \textbf{21.9} & \textbf{46.5} & \textbf{34.7} & \textbf{69.4} \\
    \midrule
    \multicolumn{11}{c}{\textbf{Inter-subject: leave one subject out for test}} \\
    \midrule
    UBP~\cite{wu2025bridging} & 2.0 & 5.7 & 1.5 & 17.2 & 2.7 & 10.5 & 2.5 & 8.0 & 2.2 & 10.4 \\
    \rowcolor{purple!20}
    \textbf{Ours} & \textbf{3.6} & \textbf{11.6} & \textbf{7.4} & \textbf{16.2} & \textbf{6.5} & \textbf{14.6} & \textbf{3.8} & \textbf{12.5} & \textbf{5.3} & \textbf{13.7} \\
    \bottomrule
\end{tabular}}
\label{tab:things-meg} 
\end{table}
\begin{figure}[t]
\centering
\includegraphics[width=\linewidth]{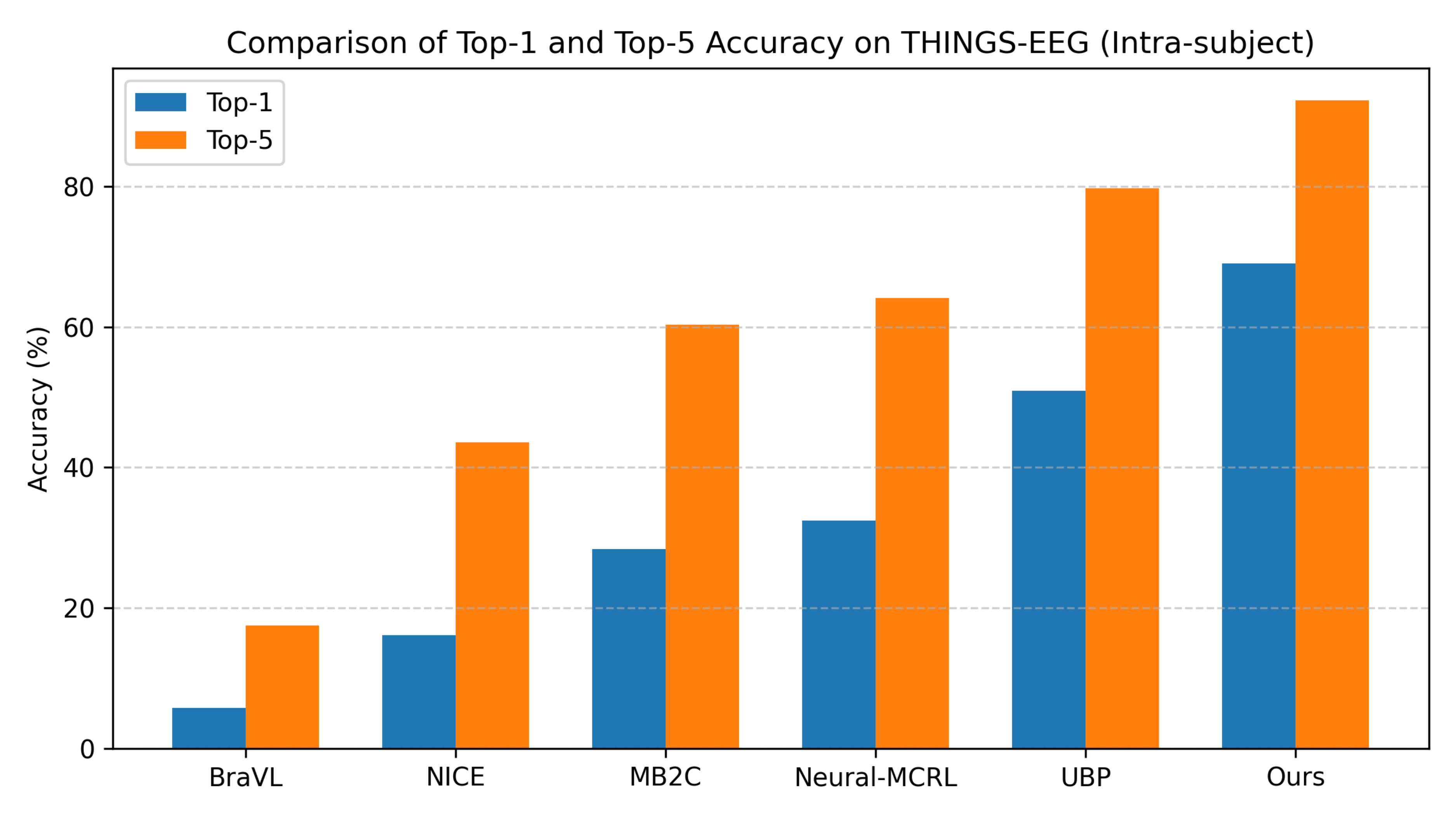}
\caption{Comparison of Top-1 and Top-5 accuracy (\%) for Intrasubject task on THINGS-EEG.}
\label{fig:performance}

\end{figure}
We compare our method with several recent representative EEG visual decoding approaches. BraVL \cite{du2023decoding} employs a mixture of experts mechanism to jointly model brain signals, visual, and linguistic features, achieving robust performance through multi-modal fusion. NICE \cite{song2024decoding} and its variants (NICE-SA, NICE-GA) introduce self-supervised EEG representation learning with different attention mechanisms for spatial modeling. MB2C \cite{wei2024mb2c} introduces multimodal bidirectional cycle consistency to learn robust EEG--visual representations. ATM-S \cite{li2024visual} focuses on adaptive temporal modeling with spatial attention for dynamic EEG encoding. CogCap \cite{zhang2025cognitioncapturer} incorporates cognitive attention-guided captioning to enhance semantic understanding. Neural-MCRL \cite{li2025neural} utilizes multi-modal contrastive representation learning to bridge neural and visual modalities. VE-SDN \cite{chen2024visual} builds a shared semantic space through visual-EEG semantic decoupling. UBP \cite{wu2025bridging}, the previous state-of-the-art, explicitly models uncertainty-aware blur prior to handle the vision-brain gap. While these methods have advanced the field, they primarily rely on global features or static semantic fusion, lacking explicit foreground-background modeling and dynamic brain network consideration, which limits their performance in complex scenes.

\paragraph{Results on THINGS-EEG}
Table~\ref{tab:things-eeg} summarizes the quantitative results on the THINGS-EEG dataset, and Figure~\ref{fig:performance} provides a visual comparison of Top-1 and Top-5 accuracy in the intra-subject setting. Our method achieves Top-1 and Top-5 accuracies of 69\% and 92.2\% on the 200-way zero-shot retrieval task, outperforming all baseline methods. In particular, it surpasses the previous state-of-the-art UBP \cite{wu2025bridging} by 18.1\% in Top-1 accuracy (69.0\% vs. 50.9\%). The performance improvements are consistently observed across all subjects, indicating stable gains.

In the inter-subject setting (leave-one-subject-out), the performance drops to 15.0\% Top-1 accuracy, with a 2.6\% improvement over UBP. Although the margin is smaller than that in the intra-subject case, the proposed method still maintains an advantage over existing approaches.

\paragraph{Results on THINGS-MEG}
Table~\ref{tab:things-meg} presents the results on the THINGS-MEG dataset. Our method achieves 34.7\% Top-1 and 69.4\% Top-5 accuracy in the intra-subject setting, outperforming UBP \cite{wu2025bridging} by 8.0\% and 14.2\%, respectively. Similar to the EEG results, consistent improvements over strong baselines are observed.

\subsection{Discussion on Cross-Subject and Cross-Modal Generalization.}

While FSDBN achieves substantial improvements in the intra-subject setting (69.0\% Top-1 on THINGS-EEG), the gains become less pronounced in cross-subject and MEG scenarios. For example, in the cross-subject setting, the Top-1 accuracy decreases to 15.0\%, and the improvement over UBP is reduced to 2.6\%. On THINGS-MEG, the Top-1 accuracy reaches 34.7\%, also with a smaller margin compared to the EEG results. These observations indicate that the effectiveness of the proposed modeling is influenced by changes in the underlying data distributions.

One possible explanation is the variability inherent in neural data. In cross-subject settings, differences in spatial organization and functional connectivity across individuals introduce shifts in the learned representations, making it more difficult to achieve reliable transfer. In cross-modal scenarios, discrepancies in sensor configuration and signal characteristics between EEG and MEG further affect the alignment between brain signals and visual features. As a result, although the proposed framework consistently improves performance, its relative advantage becomes less prominent under these conditions. Extending the current framework to better handle such variability remains an important direction for future work.

\subsection{Ablation Study}
\paragraph{Core Component Analysis}
\begin{table}[t]
\caption{Ablation study of FSDBN on THINGS-EEG.}
\label{tab:ablation}
\centering
\begin{tabular}{ccc|cc}
\hline
SCSA & SP-DGFF & DBN & Top-1 & Top-5 \\
\hline
$\times$ & $\times$ & $\times$ & 50.9 & 79.7 \\
\checkmark & $\times$ & $\times$ & 62.5 & 90.2 \\
$\times$ & \checkmark & $\times$ & 53.5 & 80.1 \\
$\times$ & $\times$ & \checkmark & 55.9 & 86.6 \\
\checkmark & \checkmark & $\times$ & 66.4 & 91.5 \\
\checkmark & $\times$ & \checkmark & \underline{66.7} & \underline{91.0} \\
$\times$ & \checkmark & \checkmark & 60.7 & 89.8 \\
\rowcolor{purple!15}
\checkmark & \checkmark & \checkmark & \textbf{69.0} & \textbf{92.2} \\
\hline
\end{tabular}
\end{table}
To evaluate the contribution of each component, we conduct ablation experiments on THINGS-EEG (Table~\ref{tab:ablation}). The baseline model achieves 50.9\% / 79.7\%. Introducing \textbf{SCSA} alone significantly improves performance to 62.5\% / 90.2\%, indicating that semantically guided foreground–background disentanglement plays a dominant role in reducing ambiguity and suppressing background interference. In contrast, \textbf{SP-DGFF} and \textbf{DBN} provide relatively limited gains when applied individually (53.5\% / 80.1\% and 55.9\% / 86.6\%), suggesting that both are less effective without reliable foreground representations. When combining two components, performance is further improved: \textbf{SCSA + DBN} (66.7\% / 91.0\%) and \textbf{SCSA + SP-DGFF} (66.4\% / 91.5\%) consistently outperform \textbf{SP-DGFF + DBN} (60.7\% / 89.8\%), highlighting the critical role of SCSA in enabling effective cross-modal alignment. Finally, the full model achieves the best performance (69.0\% / 92.2\%), demonstrating that the three components are complementary for robust EEG–visual alignment.

\begin{table}[t]
\caption{Performance with different saliency detectors.}
\label{tab:saliency}
\centering
\begin{tabular}{lcc}
\hline
Saliency Model & Top-1 (\%) & Top-5 (\%) \\
\hline
Itti-Koch & 64.2 & 90.8 \\
U2Net & \underline{66.1} & \underline{91.6} \\
\rowcolor{purple!15}
SAM(Ours) & \textbf{69.0} & \textbf{92.2} \\
\hline
\end{tabular}
\end{table}
\begin{figure}[t]
\centering
\includegraphics[width=\linewidth]{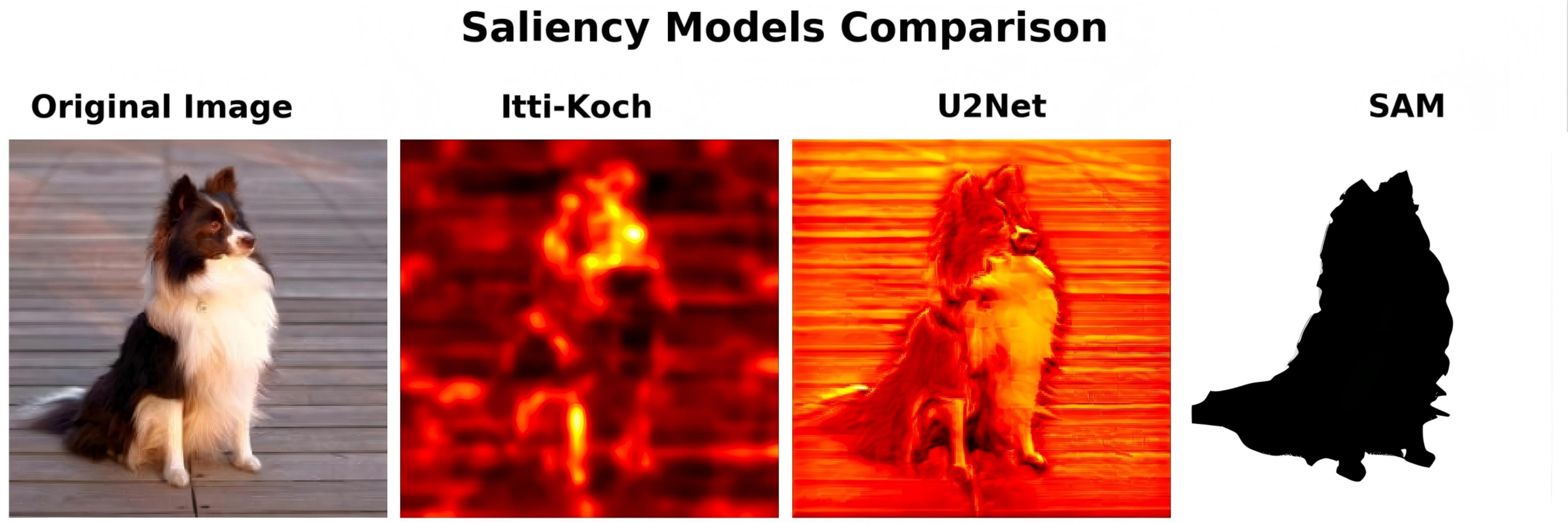}
\caption{Saliency Detection Models Comparison on Dog Image}
\label{fig:saliency}

\end{figure}
To further investigate whether SCSA relies on strong pretrained saliency models, we replace SAM with a classical model (Itti-Koch) and a lightweight deep model (U2Net). As shown in Table~\ref{tab:saliency}, all variants achieve competitive performance, with only moderate degradation compared to SAM. As illustrated in Figure~\ref{fig:saliency}, Itti-Koch produces coarse and scattered responses, while U2Net captures more complete object structures but still introduces background noise and imprecise boundaries, whereas SAM yields clean and accurate masks. Despite these substantial differences in saliency quality, SCSA maintains stable performance across all settings, indicating that its effectiveness does not depend on the strength of the saliency detector but rather on the semantic consistency constraint, which compensates for imperfect saliency maps and ensures robust foreground extraction.
\paragraph{Contribution of Saliency and Semantic Consistency}
\begin{table}[t]
\caption{Contribution analysis of saliency and semantic consistency.}
\label{tab:scsa}
\centering
\begin{tabular}{lcc}
\hline
Variant & Top-1 (\%) & Top-5 (\%) \\
\hline
w/o semantic consistency & \underline{63.8} & \underline{90.9} \\
w/o saliency map & 61.7 & 89.5 \\
\rowcolor{purple!15}
SCSA(Ours) & \textbf{69.0} & \textbf{92.2} \\
\hline
\end{tabular}
\end{table}

We further analyze the contribution of saliency and semantic consistency, as shown in Table~\ref{tab:scsa}. Removing the semantic consistency loss results in a clear performance drop to 63.8\% / 90.9\% (-5.2\% Top-1), indicating that high-level semantic alignment plays a critical role in guiding the model toward semantically meaningful regions. Removing the saliency map leads to an even larger degradation (61.7\% / 89.5\%, -7.3\% Top-1), suggesting that explicit spatial priors are essential for effective foreground–background separation. Notably, the full SCSA achieves the best performance (69.0\% / 92.2\%), demonstrating that neither saliency nor semantic consistency alone is sufficient. Instead, their combination provides complementary supervision, where saliency offers spatial localization while semantic consistency enforces high-level alignment, jointly enabling robust and semantically grounded visual representation learning.
\paragraph{Effectiveness of Structured FG/BG Decomposition}
\begin{table}[t]
\caption{Effectiveness of structured foreground-background decomposition.}
\label{tab:fgbg}
\centering
\begin{tabular}{lcc}
\hline
Method & Top-1 (\%) & Top-5 (\%) \\
\hline
Random FG/BG & 60.3 & 88.9 \\
\rowcolor{purple!15}
 SP-DGFF(Ours) & \textbf{69.0} & \textbf{92.2} \\
\hline
\end{tabular}
\end{table}

Finally, we evaluate whether the foreground–background decomposition captures meaningful structure or merely acts as a form of random augmentation. As shown in Table~\ref{tab:fgbg}, replacing the proposed structured decomposition with random partitioning leads to a significant performance drop to 60.3\% / 88.9\% (-8.7\% Top-1), indicating that naive foreground-background separation fails to preserve semantic relevance. In contrast, our SP-DGFF explicitly models semantically guided foreground and background features, enabling more effective feature disentanglement and interaction. These results demonstrate that the proposed decomposition captures meaningful semantic structure rather than introducing random perturbations, and plays a critical role in improving cross-modal alignment.

\subsection{Semantic Alignment Analysis and Visualization}
\begin{figure}[t]
\centering
\includegraphics[width=\linewidth]{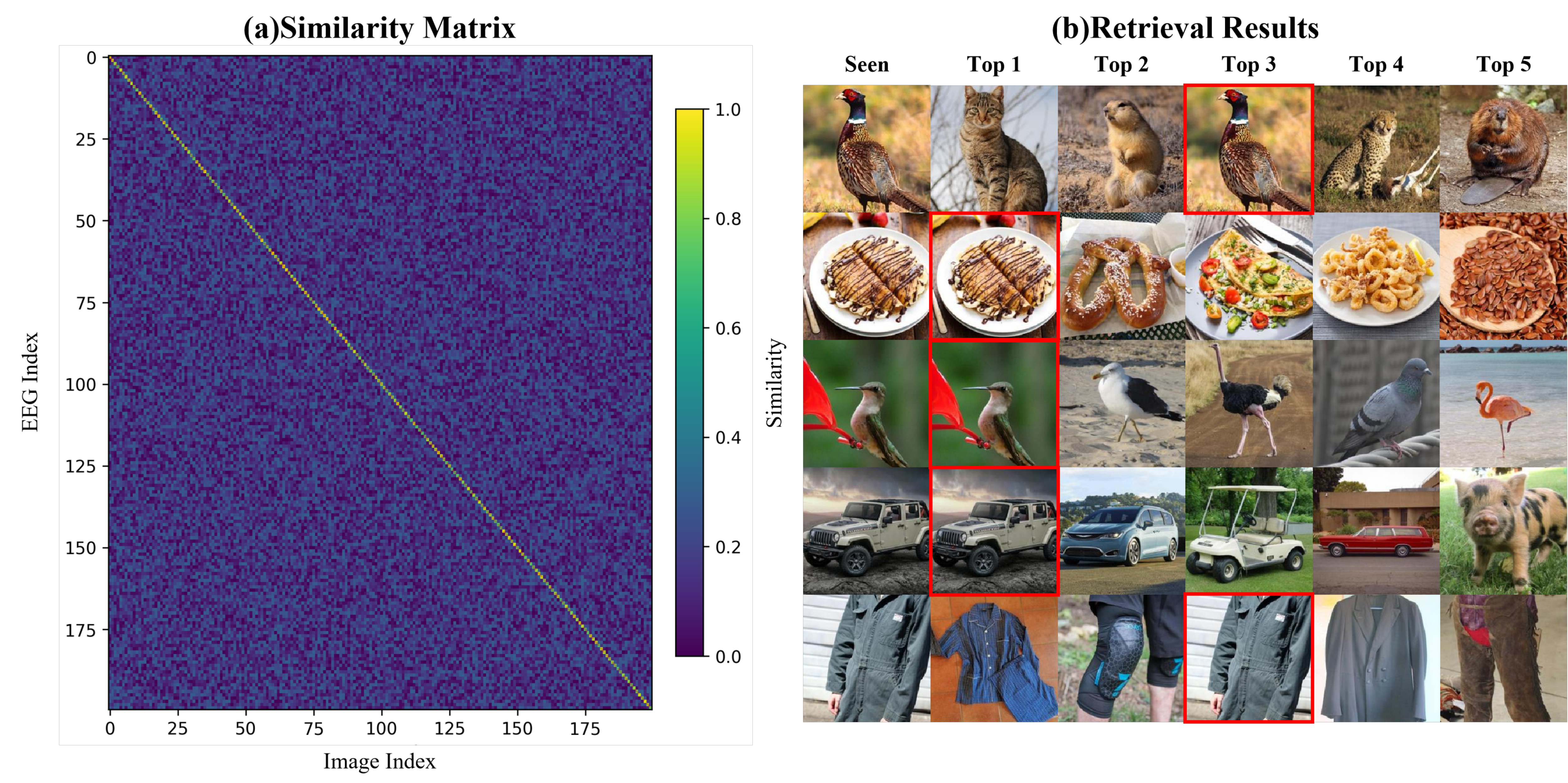}
\caption{Left: Similarity matrix between EEG queries and image candidates. Right: Qualitative retrieval results. Red boxes indicate ground-truth matches.}
\label{fig:similarity}

\end{figure}
To validate the geometric structure of the learned cross-modal embedding space, we visualize the pairwise similarity matrix between EEG queries and image candidates (Figure~\ref{fig:similarity}, left). The clear diagonal dominance indicates that EEG representations align well with their corresponding visual stimuli, while the localized block patterns further suggest that the model captures implicit semantic groupings across categories. This structured organization demonstrates that the Dynamic Brain Network effectively models category-related neural patterns, and the SCSA module aligns these patterns with semantically relevant foreground regions.

Qualitative retrieval results (Figure~\ref{fig:similarity}, right) further support this observation. The top-$k$ candidates are consistently semantically coherent, even when the exact match is not ranked first, and no obvious cross-category errors are observed. Moreover, meaningful intra-class ranking (e.g., visually similar instances ranked higher) indicates that the model captures fine-grained semantic distinctions. These results suggest that the semantic-driven gating mechanism effectively suppresses background interference and enables more discriminative cross-modal representations.

\subsection{Similarity Score Distribution Analysis}
\begin{figure}[t]
\centering
\includegraphics[width=\linewidth]{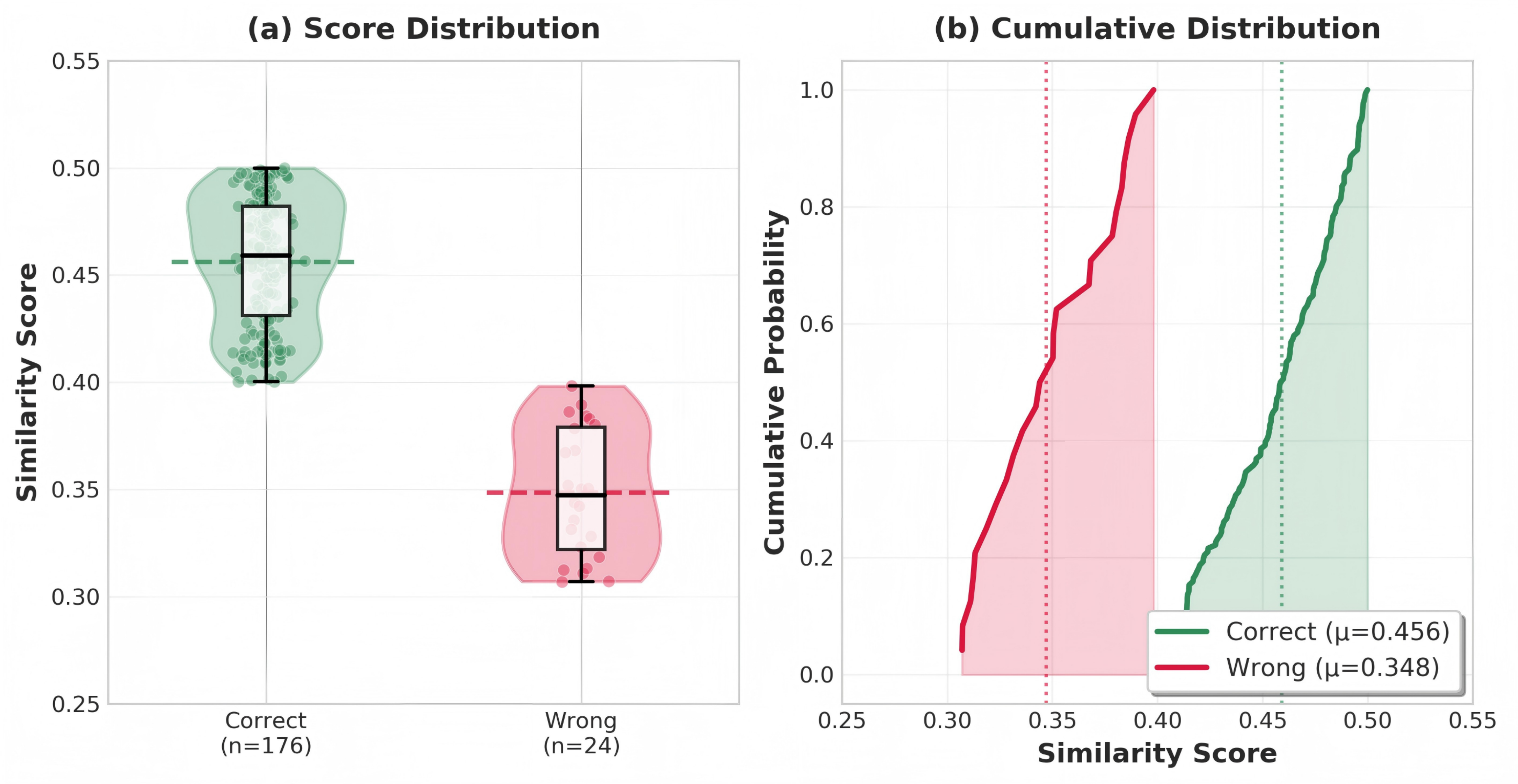}
\caption{(a) Score distribution for correct and wrong retrievals. (b) Cumulative distribution function comparing the two groups.}
\label{fig:distribution}

\end{figure}
To further analyze the discriminative capability of our learned embeddings, we examine the distribution of similarity scores between EEG queries and retrieved images. As shown in Figure~\ref{fig:distribution}(a), correct retrievals ($n=176$) exhibit significantly higher similarity scores ($\mu=0.456$, $\sigma=0.037$) compared to wrong retrievals ($n=24$, $\mu=0.348$, $\sigma=0.037$). The pronounced separation ($p<0.001$, two-sample $t$-test; Cohen's $d=2.92$) validates the effectiveness of our embeddings. Notably, Figure~\ref{fig:distribution}(b) reveals that $80\%$ of correct matches achieve scores above $0.45$, whereas $80\%$ of wrong matches fall below $0.35$, indicating strong separability. This substantial gap of $0.108$ in mean scores demonstrates that our dynamic brain network modeling successfully encodes discriminative neural signatures for reliable visual decoding.
\section{Conclusion}
In this paper, we propose FSDBN, a neuro-inspired framework that bridges the gap between static visual features and dynamic neural processes. FSDBN explicitly decouples semantic foregrounds from background clutter rather than treat scenes holistically, and aligns them with adaptive spatiotemporal brain networks. By integrating semantic-consistent saliency with dynamic gating, our approach effectively suppresses background interference while capturing the time-varying neural reorganization driven by focal attention. Extensive experiments establish a new state-of-the-art on THINGS-EEG and THINGS-MEG, demonstrating the necessity of modeling the dynamic interaction between visual saliency and brain connectivity. Future work will focus on optimizing computational efficiency, improving cross-subject generalization, and exploring lightweight architectures for real-time applications.

\balance

\bibliographystyle{unsrtnat}
\bibliography{fsdbn_arxiv}

\end{document}